%% file: 0000_MAIN_general.tex
\documentclass[a4paper]{article}
\usepackage{multirow}
\usepackage{graphicx}
\usepackage{float}
\usepackage{dcolumn}
\usepackage{bm}
\usepackage{booktabs}
\usepackage{subfig}
\usepackage{caption}
\usepackage{graphicx}
\usepackage{amsmath}
\usepackage{amsfonts}
\usepackage{amssymb}
\usepackage{amsthm}
\usepackage{authblk}
\usepackage{physics}
\usepackage{algorithms/algorithm}
\usepackage{algpseudocode}
\usepackage[export]{adjustbox}
\usepackage[margin=1.5 in]{geometry}
\providecommand{\keywords}[1]
{
  \small	
  \textbf{\textit{Keywords---}} #1
}
\begin{document}


\title{A Quantum Approximate Optimization Method For Finding Hadamard Matrices}

\author{Andriyan Bayu Suksmono}
\affil{
The School of Electrical Engineering and Informatics\\
Institut Teknologi Bandung, Indonesia 
}


\date{\today}
\maketitle

\input{0x_Abstract}


\input{1__Introduction}
%
\input{2__Results}
%
\input{3__Discussions}

%
\input{4__Methods.tex}


\bibliographystyle{unsrt}
\bibliography{0000_MAIN_general}

\input{5_Acknowledgements.tex}

%
\clearpage
%
\end{document}

%% file: 0x_Abstract.tex
\begin{abstract}
Finding a Hadamard matrix of a specific order using a quantum computer can lead to a demonstration of practical quantum advantage. Earlier efforts using a quantum annealer were impeded by the limitations of the present quantum resource and its capability to implement high order interaction terms, which for an $M$-order matrix will grow by $O(M^2)$. In this paper, we propose a novel qubit-efficient method by implementing the Hadamard matrix searching algorithm on a gate-based quantum computer. We achieve this by  employing the Quantum Approximate Optimization Algorithm (QAOA). Since high order interaction terms that are implemented on a gate-based quantum computer do not need ancillary qubits, the proposed method reduces the required number of qubits into $O(M)$. We present the formulation of the method, construction of corresponding quantum circuits, and experiment results in both a quantum simulator and a real gate-based quantum computer.
\end{abstract}

%% file: 1__Introduction.tex
\section{Introduction}
Quantum computing achieved a significant milestone in 2019 when Google's quantum computer, Sycamore, outperformed a classical supercomputer in a specialized task known as random quantum circuit sampling \cite{arute_2019}. While a classical super computer needed about 10,000 years, the 53 qubits Sycamore took around 200 seconds to finish the task, thanks to its capability in representing $2^{53} \approx 10^{16}$ computational state-space. The next stage after this milestone, according to this paper, is showing the capability of a quantum computer to solve a more valuable computing applications. Although an ideal fault-tolerant system with a sufficient number of qubits for implementing practical quantum algorithms has not yet been achieved—placing us in what is known as the NISQ (Noisy Intermediate-Scale Quantum) era—various efforts in this direction have already been initiated. One potential approach for using NISQ devices to solve real-world computing problems is through hybrid classical-quantum algorithms, such as the Quantum Approximate Optimization Algorithm (QAOA) proposed by Farhi et al. \cite{farhi_2014}.

To date, researchers have conducted extensive theoretical studies, enhancements, and explorations of potential applications of the QAOA. In \cite{boulebnane_2023}, Boulebnane et al. reported their investigation on the performance of QAOA in sampling low-energy states for protein folding problems. Their results indicate that, whereas simpler problems give promising results, a more complex one that required a deeper quantum circuit only comparable to that of random sampling. He et al. \cite{he_2023} investigated the selection of the initial state in the Quantum Approximate Optimization Algorithm (QAOA) for a constrained portfolio optimization problem. By leveraging the close relationship between QAOA and the adiabatic algorithm, they found that the optimal initial state corresponds to the ground state of the mixing Hamiltonian. Improvement to the QAOA performance is also actively being explored. A double adaptive-region Bayesian optimization for QAOA which indicates a better performance in terms of speed, accuracy, and stability, compared to conventional optimizer is reported by Cheng et al. \cite{cheng_2024}. On the application side, a data-driven QAOA for distributed energy resource problem in power systems is reported by Jing et al. \cite{jing_2023}. 

Another significant result in the use of NISQ devices is the recent demonstration of quantum utility before fault tolerance by IBM researchers \cite{kim_2023}. This result brings hopes on the implementation and demonstration of quantum advantage for real-world applications. In line with this spirit, we propose a hard problem of discovering a particular discrete structure--which is a specific order of Hadamard matrix, as a potential instance of such practical applications and use QAOA for implementation in gate-based quantum computers. 

A Hadamard matrix (H-matrix) is an orthogonal binary matrices with various scientific and engineering applications \cite{hadamard_1893, sylvester_1867, garg_2007, horadam_2007}. An $M$-order H-matrix exists only when $M$ equal to 1, 2, and multiples of 4. The converse, that for every positive integer $k$ there is a Hadamard matrix of order $4k$ is also believed to be true \cite{hedayat_1978, horadam_2007}, which is the well known Hadamard matrix conjecture. When $M=2^n$, for a non-negative integer $n$, the H-matrix can be constructed easily by Sylvester method \cite{hedayat_1978}. Construction of H-matrix with other values of $M=4k$ has also been developed, among others are the methods by Paley \cite{paley_1933}, Williamson \cite{williamson_1944}, Baumert-Hall \cite{baumert_hall_1965}, and Turyn \cite{turyn_1974}. More recently, co-cylic techniques are developed by Delauney-Horadam \cite{horadam_1993, horadam_1995, horadam_2000}, and Alvarez et al. \cite{alvarez_2020}. Nevertheless, not all of Hadamard matrices are neither easily constructed nor discovered. The latest one is a H-matrix of order 428, which was found by Kharaghani and Tayfeh-Rezaie \cite{kharaghani_rezaie_2004}. Up to this day, for order $M<1000$, the H-matrices of order $668, 716$, and $892$ have neither been discovered nor proven to exist. Our previous study indicates that, by using currently known methods, present-day (classical) computing resources are insufficient to find those matrices in practical time.

In principle, an $M$-order H-matrix can be found, or proven non-existent, by exhaustively testing all possible $+1/-1$ combinations of its $M \times M$ elements. However, when $M$ becomes sufficiently large, this approach becomes computationally impractical, as the number of orthogonality tests grows exponentially, by $O(2^{M \times M})$, even though each test can be performed in polynomial time. To address this issue, we have developed several methods based on Simulated Annealing (SA) and Simulated Quantum Annealing (SQA) \cite{suksmono_2018}, and Quantum Annealing (QA) \cite{suksmono_minato_2019, suksmono_2022}. Our latest method was implemented on a quantum annealer, specifically the D-Wave quantum computer, where we successfully discovered several H-matrices up to order greater than one hundred \cite{suksmono_2022}. Although current quantum annealers have over 5,000 qubits, the need for ancillary qubits to represent interactions beyond 2-body limits hinders implementation to find higher-order H-matrices. We have estimated that the implementation of the method for finding a $668$-order H-matrix needs at least $15,400$ physical qubits \cite{suksmono_2022}. 

A tentative way to pursue this task is by developing a qubit-efficient method.  This paper explores this idea by proposing the use of a circuit/gate-based quantum computer instead of a quantum annealer to implement the method. An almost straightforward extension for the previous method is by formulating the problem as an instance of the QAOA (Quantum Approximate Optimization Algorithm) method \cite{farhi_2014}. In a gate-based quantum computer, the number of interaction in the Hamiltonian terms is not limited to only the 2-body interaction,  unlike in quantum annealers. The additional ancillary qubits required for implementing higher-order interaction terms are unnecessary when using a gate-based quantum computer. This approach significantly reduces the total number of qubits required.

%% file: 2__Results.tex
\section{Results}
In this paper, we use two types of binary variables: Boolean variables $b$, which take values of either 0 or 1, and spin variables $s$, which take values of either -1 or +1. A Boolean value of 0 is mapped to +1 in the spin variable, and vice versa, while a Boolean value of 1 is mapped to -1 in the spin variable, and vice versa. For example, a Boolean vector $\vec{b}=[b_0,b_1,b_2,b_3,b_4]^T= [0,1,0,1,1,0]^T$--simply written as a bit string   $\vec{b}=b_0b_1b_2b_3b_4= 010110$; is mapped to the spin vector $\vec{s}=[s_0,s_1,s_2,s_3,s_4]^T
=[1, -1, 1, -1, -1, 1]^T$, where $(\cdot)^T$ denote transpose. The Boolean and spin variables will be used interchangeably, depending on the context.
\input{2_1_QAOA_4_HSEARCH}
\input{2_2_Performance_Metric}
\input{2_3__Experiments}

%% file: 2_1_QAOA_4_HSEARCH.tex
\subsection{QAOA Formulation of H-matrix Searching Problem}

The QAOA is a hybrid classical-quantum algorithm proposed by Farhi et al. \cite{farhi_2014}.  It is proposed as a solution for near-term quantum computing, which can be implemented on a Noisy Intermediate-Scale Quantum (NISQ) devices—quantum computers with limited qubits, connectivity, gate errors, and short coherence times. A typical $N$-bit and $M$-clause combinatorial optimization problem addressed by the QAOA can be formulated as follows.  Consider an $N$-length bit string $\vec{b}=b_0b_1 \cdots b_{N-1}$ and let $C(\vec{b})$ be a cost or an objective function given by the following expression
\begin{equation}
    C(\vec{b}) = \sum_{m=0}^{M-1} C_m(\vec{b})
    \label{EQ_QAOA_basic}
\end{equation}
The value of $C_m(\vec{b})$ is equal to 1 if $\vec{b}$ satisfies the clause $C_m$, otherwise it is $0$ . When $C$ is the maximum value of Eq. (\ref{EQ_QAOA_basic}), the approximation means that we seek for a bit string $\vec{b}$ where $C(\vec{b})$ is close to $C$.

For applying the QAOA to the H-matrix searching problem, we change the Boolean vector $\vec{b}$ in Eq. (\ref{EQ_QAOA_basic}) into its spin vector representation $\vec{s}=[s_0, s_1, \cdots, s_M]$, while the maximization is recast as minimization. We can restate the previous approximation problem into finding a vector $\vec{s}$ that minimize a non-negative cost function given by

\begin{equation}
    C(\vec{s}) = \sum_{m=0}^{M-1} C_m(\vec{s})
    \label{EQ_QAOA_spin}
\end{equation}
Then, the approximation means that we seek for a bit string $\vec{b}$ corresponding to the vector $\vec{s}$ that makes $C(\vec{s})$ close to zero. The bit string $\vec{b}$ is obtained by measuring the qubits after executing the QAOA in a circuit-based quantum computer, which is done by evolving the system under a Hamiltonian $\hat{H}$ that represents a given problem.

In the QAOA method, we have a Hamiltonian $\hat{H}$ that consists of a \emph{problem Hamiltonian} $\hat{H}_C$ and a \emph{mixer Hamiltonian} $\hat{H}_B$,
\begin{equation}
    \hat{H} = \hat{H}_C + \hat{H}_B
    \label{EQ_HC_HB}
\end{equation}
The problem Hamiltonian $\hat{H}_C$ is equivalent to the potential Hamiltonian $\hat{H}_{pot}$ in the quantum annealing formulation \cite{suksmono_2022}. It is derived from the cost function of the Hadamard searching problem, either in the Williamson/Baumert-Hall (WBH) method $C_W(\vec{s})$ or the Turyn method $C_T(\vec{s})$. The details of these cost functions are provided in the Methods section. 

Based on Eq. (\ref{EQ_HC_HB}), we construct a quantum circuit to perform the following unitary transform  
\begin{equation}
      \hat{U}(\gamma, \beta) = e^{-i\beta_P \hat{H}_B}e^{-i\gamma_P \hat{H}_C} e^{-i\beta_{P-1} \hat{H}_B}e^{-i\gamma_{P-1} \hat{H}_C} \cdots e^{-i\beta_p \hat{H}_B}e^{-i\gamma_p \hat{H}_C}  \cdots e^{-i\beta_1 \hat{H}_B}e^{-i\gamma_1 \hat{H}_C}
    \label{EQ_QAOA_general}%
\end{equation}
where
\begin{equation}
  \hat{H}_B=\sum_j b_j \hat{\sigma}^x_j  
  \label{EQ_QAOA_driver}%
\end{equation}
and
\begin{equation}
    \hat{H}_C=\sum_{j,k, \cdots,m,n} c_{jk\cdots mn} \hat{\sigma}^z_j\hat{\sigma}^z_k\cdots \hat{\sigma}^z_m\hat{\sigma}^z_n  
  \label{EQ_QAOA_problem}%
\end{equation}
In these equations, $P$ is the number of layers (Trotter slice),  $\gamma_p$ and $\beta_p$ are  (angle) parameters at layer $p$, $b_j$ and $c_{j,k,\cdots, m,n}$ are constants, whereas $\hat{\sigma}^x_j$ and $\hat{\sigma}^z_j$ are the $j^{th}$ spin/Pauli matrices in $x$ and $z$-directions, respectively.

\begin{figure}[h!]
    \centering
    \includegraphics[width=1.0\columnwidth] {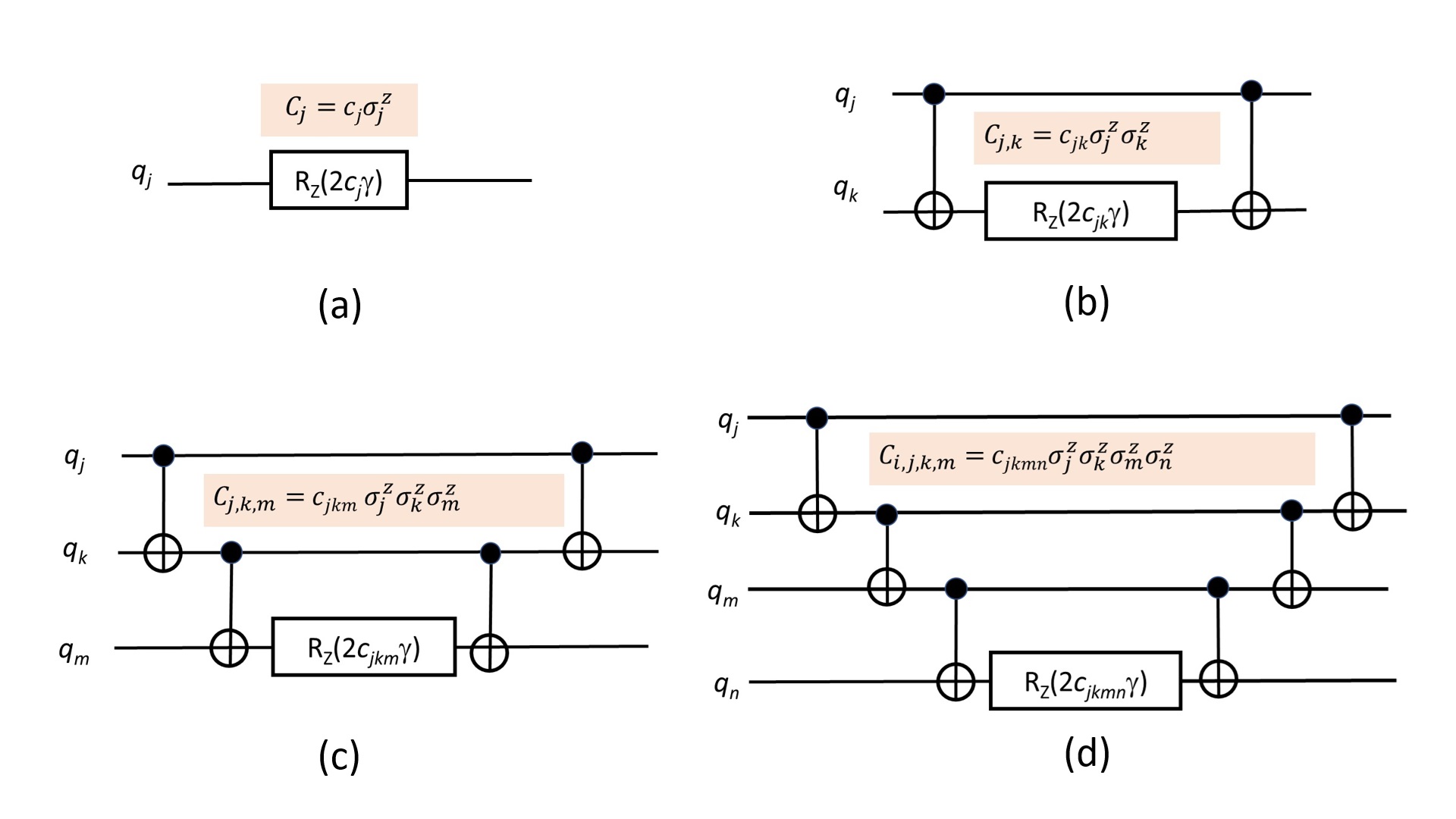} 
    \caption{Elementary quantum circuits used in the QAOA-based Hadamard matrix search, categorized into four types: (a) 1-body interaction term, (b) 2-body interaction term, (c) 3-body interaction term, and (d) 4-body interaction term. In this figure, $q_j, q_k, q_m,$ and $q_n$ are the $j^{th}, k^{th}, m^{th},$ and $n^{th}$ qubits, respectively, $\gamma$ denotes rotation angle and $c_j, c_{jk}, c_{jkm}, $ and $c_{jkmn}$ are constants.}
    \label{FIG_1BODY}
\end{figure}
The term expressed by the product of $n$ Pauli matrices $\hat{\sigma}_0^z\hat{\sigma}_1^z \cdots \hat{\sigma}_{n-1}^z$ in Eq.(\ref{EQ_QAOA_problem}) is called an $n$-body interaction term. In the Hadamard Searching Problem (H-SEARCH), there are only up to 4-body interaction in the Hamiltonian, so that generally $\hat{H}_C$ can be expressed by

\begin{equation}
    \hat{H}_C=\sum_{j} c_j \hat{\sigma}^z_j  +
    \sum_{j,k}c_{jk}\hat{\sigma}^z_j\hat{\sigma}^z_k +
    \sum_{j,k,m} c_{jkm} \hat{\sigma}^z_j\hat{\sigma}^z_k\hat{\sigma}^z_m + 
    \sum_{j,k,m,n} c_{jkmn} \hat{\sigma}^z_j\hat{\sigma}^z_k\hat{\sigma}^z_m\hat{\sigma}^z_n 
  \label{EQ_QAOA_problem_hsearch}%
\end{equation}
Here, the $1^{st}$, $2^{nd}$, $3^{rd}$, and $4^{th}$ terms corresponds to the 1-body, 2-body, 3-body, and 4-body interaction terms, subsequently. The quantum circuits representing each of these elementary terms are displayed in Fig.\ref{FIG_1BODY}.

%% file: 2_2_Performance_Metric.tex
\subsection{Performance Metrics} 

The outputs of the algorithms discussed in this paper are $L$-length bit strings, resulting in $2^L$ possible combinations. An output string is considered correct or valid if the value of the associated non-negative error or energy function—such as the Williamson or Turyn cost function—is equal to zero. Otherwise, it is labeled as incorrect or wrong. To evaluate the algorithm's performance in producing correct solutions, we compare it to an algorithm that randomly generates all possible $L$-length bit strings. Accordingly, we introduce the xRAR (\emph{x-algorithm to Random-Algorithm Ratio}) as a performance metric for a given x-algorithm.

\begin{figure}[h!]
    \centering
    \includegraphics[width=1.0\columnwidth] {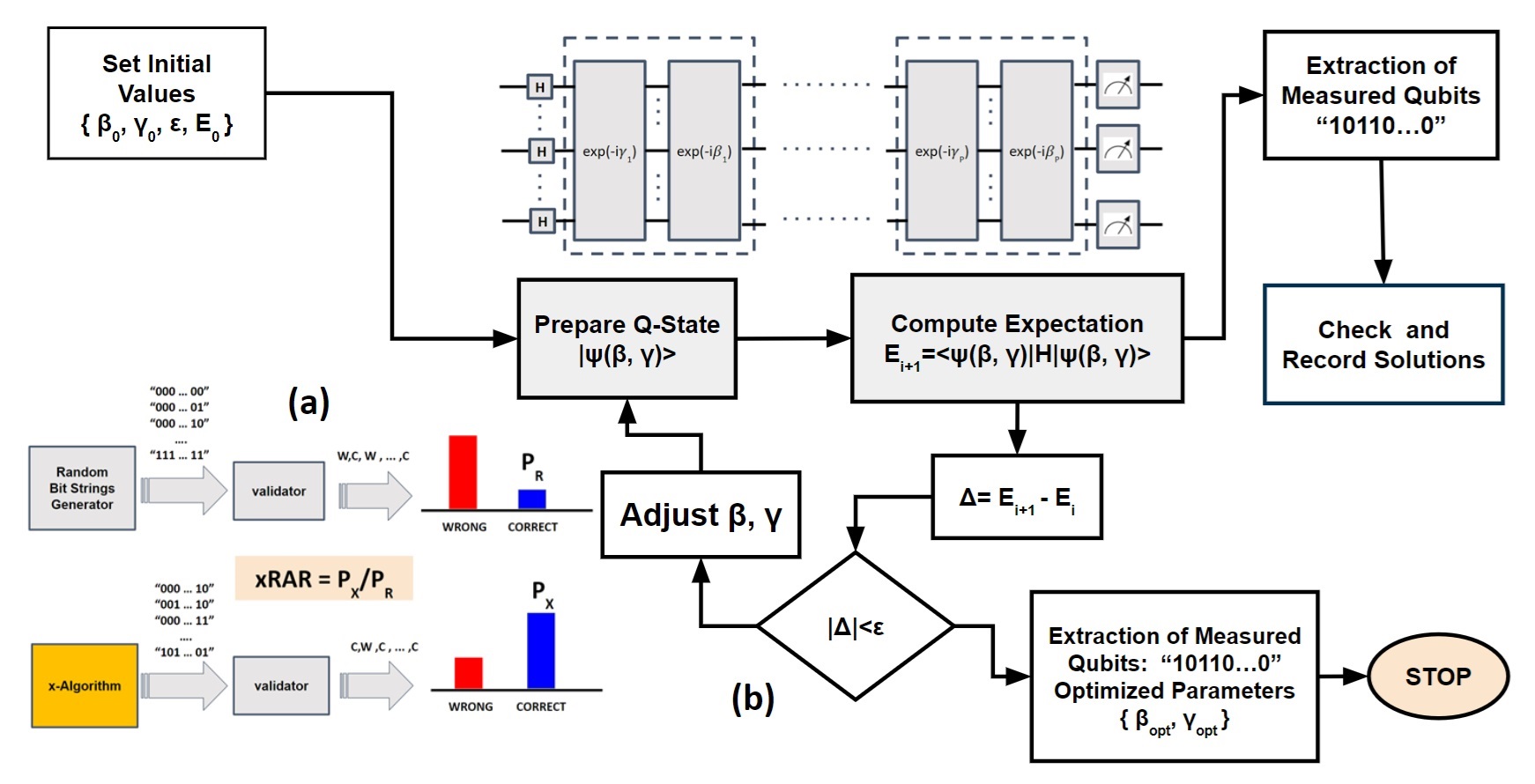}
    \caption{Performance metric and workflow diagram of the QAOA: (a) Performance measure in term of xRAR. The probability $P_x$ of correct or valid answers of the x-algorithm that generates $L$-length bit strings is compared to $P_R$, which is the correct probability of a random algorithm that generates $L$-length (uniform) randomly distributed bit strings, (b)The QAOA processing steps in the experiments. The parameters $\{\beta, \gamma\}$ are angle parameters which is updated during the execution of hybrid quantum-classical algorithm. The threshold value $\epsilon$ is a stopping parameter. When the difference $\Delta$ between current energy $E_{i+1}$ and previous one $E_{i}$ is considered insignificant, the execution of the algorithm is terminated.}
    \label{FIG_experiment_flow} 
\end{figure}

The random algorithm generates $2^L$ possible bit strings and we can evaluate whether each of the bit string is a correct/valid solution or not. If there are $S_R$ correct solutions among the $2^L$ random bit strings and we assume a uniform probability distribution among them, the probability of finding a valid solution, $P_R$, is given by $P_R = \frac{S_R}{2^L}$. Similarly, the solutions generated by a quantum circuit that represents the x-algorithm are also probabilistic, and we can calculate the probability $P_x$ of correct solution of the x-algorithm. Therefore, the value of xRAR; which conceptually is illustrated by Fig.\ref{FIG_experiment_flow} (a), is given by

\begin{equation}
    xRAR = \frac{P_x}{P_R}
    \label{EQ_XRAR}
\end{equation}
The value of xRAR in Eq. (\ref{EQ_XRAR}) is a positive real number. An xRAR value where $0 < \textit{xRAR} < 1$ indicates that the x-algorithm performs worse than the random algorithm $R$, $\textit{xRAR} = 1$ signifies comparable performance to the random algorithm, and $\textit{xRAR} > 1$ suggests that the x-algorithm outperforms the random algorithm $R$.

The execution steps of the QAOA used in the experiments are shown in Fig. \ref{FIG_experiment_flow} (b). When running the quantum algorithm, either on a simulator or a real quantum computer, we repeat the process $N$ times, referred to as the number of shots. This produces $N$ solutions or bit strings, each with a corresponding value of energy or error. A specific part of the algorithm computes the average or expectation value, allowing us to determine whether any of the bit strings achieve the minimum energy, as indicated by the \emph{Extraction of Measured Qubits} blocks in the figure. We can count the number of valid solutions at each iteration step and also after reaching the lowest average energy for a given experimental setup. The number of correct solutions is then used to evaluate the algorithm's performance.

In addition to xRAR, we also evaluate the performance of the algorithm using an error metric. This error metric is defined as the accumulated or total objective values of all generated solutions, with the objectives measured by either the Williamson or Turyn cost functions (Eq. (\ref{EQ_cost_williamson}) and Eq. (\ref{EQ_cost_turyn}) in the Methods section), respectively. The error is normalized by the maximum value of each cost function and then compared to the value obtained by a random algorithm through exhaustive search.

For a particular order of H-SEARCH that generates $N_Q$-length bit string solutions with a maximum objective error of $E_{max}$ and a total error for all possible $2^{N_Q}$ bit strings equal to $E_{tot}$, the normalization factor is $2^{N_Q}E_{max}$. The average error is given by $E_{tot}/2^{N_Q}$, and the normalized average error is $E_{tot}/(2^{N_Q} E_{max})$. This normalized average error is used to compare the performance of algorithms with varying numbers of shots (samples).

%% file: 2_3__Experiments.tex
\subsection{Experiments and Analysis}
The experiments presented in this paper were conducted using both simulators and quantum hardware. For the latter, we carried out the experiments on an IBM quantum computer. Before implementing the quantum circuit for the QAOA-based Hadamard search, which involves multiple $k$-body interaction terms, we first validate the elementary circuits shown in Fig.\ref{FIG_1BODY}, both individually and in combination. 

\subsubsection{Evaluation of Elementary Quantum Circuits}
\begin{figure*}[h!]
    \centering
    \subfloat[\centering Realized Quantum Circuit in Qiskit]
    {{\includegraphics[width=0.55\columnwidth]{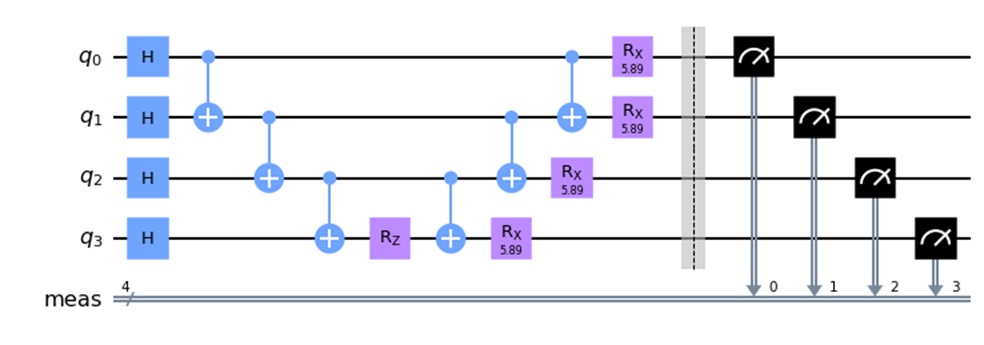} }} 
    \subfloat[\centering Distribution of solution]
    {{\includegraphics[width=0.40\columnwidth]{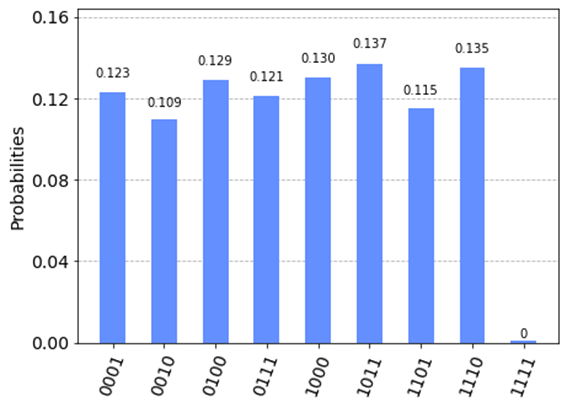} }} %
    \caption{Evaluation of the elementary circuit for the QAOA-based Hadamard matrix search: (a) A single-layer QAOA quantum circuit implemented for a problem involving a 4-body interaction term, (b) The distribution of solutions after running the circuit on a quantum computer simulator. The resulting histogram demonstrates that all solutions are found with an almost uniform distribution}%
    \label{FIG_4BODYX}%
\end{figure*}

The prototypical problem that we use to evaluate the individual elementary circuits is finding matrix's element(s) of a 2-order H-matrix. For cases involving the 1,2,3, and 4 unknown elements, the representative problem in finding unknown elements $s_1, s_2, s_3, s_4$ are given by the following $2\times 2$ matrices
\begin{equation}
    \nonumber
    \begin{pmatrix}
        1 & 1  \\
        1 & s_1
    \end{pmatrix},
    \begin{pmatrix}
        1 & s_1  \\
        1 & s_2
    \end{pmatrix},
    \begin{pmatrix}
        1 & s_1  \\
        s_2 & s_3
    \end{pmatrix},
    \begin{pmatrix}
        s_1 & s_3  \\
        s_2 & s_4
    \end{pmatrix}
\end{equation}
Then, for each cases, we formulate their corresponding energy functions $E(s)$ and Hamiltonians $\hat{H}(\hat{\sigma})$, whose formulation has been described in our previous papers \cite{suksmono_minato_2019, suksmono_2022}. As an example, for 4 unknown elements of the latest matrix, the Hamiltonian is
\begin{equation}
    \nonumber
    \hat{H}\left( \hat{\sigma}\right)=2\left(1+\hat{\sigma}^z_{1}\hat{\sigma}^z_{2} \hat{\sigma}^z_{3}\hat{\sigma}^z_{4} \right) 
    \label{EQ_4body_hsearch_hamiltonian}
\end{equation}
This Hamiltonian includes a 4-body interaction term, with its corresponding quantum circuit shown in Fig. \ref{FIG_1BODY}(d). After incorporating the mixing Hamiltonian, the quantum circuit for this QAOA problem is shown in Fig. \ref{FIG_4BODYX}(a). The solution distribution obtained from execution on a quantum simulator is displayed in Fig. \ref{FIG_4BODYX}(b).

This case represents a simple, non-cascaded QAOA problem that can be effectively solved using a single-layer quantum circuit. The solutions with dominant probabilities, as shown in the histogram—{0001, 0010, 0100, 0111, 1000, 1011, 1101, 1110}—are all correct. The total probability of these solutions results in a performance value of $xRAR = 1.998$, which is nearly the maximum possible value of $xRAR = 2$ for this problem. 

However, when elementary circuits are combined into a more complex circuit, the problem becomes increasingly difficult to solve. For a general problem involving mixed $k$-body cascaded circuits, we consider two cases for the Hamiltonians: uniform coefficients and non-uniform coefficients.

\begin{figure*}[h!]
    \centering
    \subfloat[\centering Distribution of solution for $p=1$]
    {{\includegraphics[width=0.45\columnwidth]{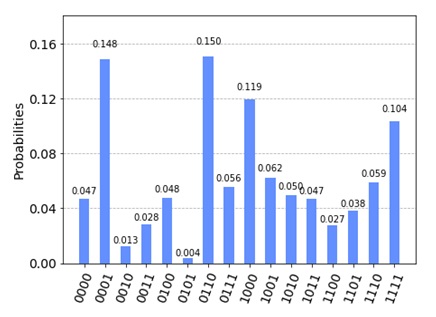} }} 
    \subfloat[\centering Distribution of solution for $p=32$]
    {{\includegraphics[width=0.45\columnwidth]{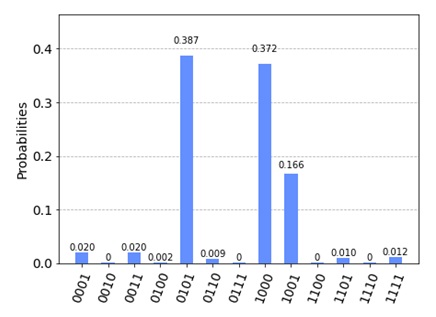} }} 
    \\
    \subfloat[\centering Iteration curve for $p=32$]
    {{\includegraphics[width=0.48\columnwidth,valign=t]{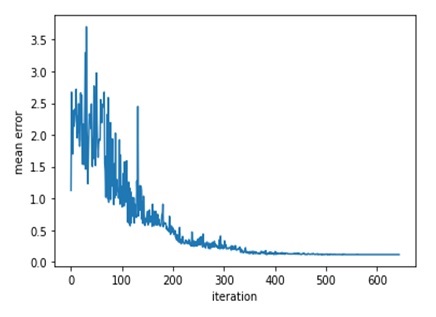} }}%
    \subfloat[\centering Table of performance]
    {{ \begin{tabular}[t]{|c|c|r|r|}
            \hline
            No & N-layer $p$ & Obj. Error&    xRAR\\
            \hline
            1&  1&   1.850&   1.344\\
            \hline
            2&  4&   0.945&   2.117\\
            \hline
            3&  8&   0.337&   3.313\\
            \hline
            4&  16&   0.465&   3.121\\
            \hline
            5&  32&   0.115&   3.734\\
            \hline
       \end{tabular}
     \vphantom{\includegraphics[width=0.48\columnwidth,valign=t]{img/k_bodies_uniform_coeffs_iter_p_2NQ_sqr.jpg}}
     }}
    \caption{Experiment results for a mixed k-body Hamiltonian with uniform coefficients: (a) distribution of solution for $p=1$ showing both of correct and incorrect solutions with with comparable probability values, (b) distribution of solution for $p=32$, indicating that increasing the layer number improve the solution; non correct solutions are suppressed, (c) Curve showing decreasing of error by iteration and achieve convergence after more than 500 iteration, (d) performance table showing that increasing the number of layer will improve the performance, i.e, reducing objective error and increasing the xRAR.}%
    \label{FIG_kbody_uniform_coeffs}%
\end{figure*}

First, we consider a Hamiltonian with uniform coefficients for all terms, expressed as follows:
\begin{equation}
    \hat{H}\left( \hat{\sigma}\right)= \hat{\sigma}^z_{0} + \hat{\sigma}^z_{1} + \hat{\sigma}^z_{0}\hat{\sigma}^z_{1} + \hat{\sigma}^z_{1}\hat{\sigma}^z_{2} + \hat{\sigma}^z_{1}\hat{\sigma}^z_{2} \hat{\sigma}^z_{3} + \hat{\sigma}^z_{0}\hat{\sigma}^z_{1}\hat{\sigma}^z_{2} \hat{\sigma}^z_{3}-1
    \label{EQ_cascaded_uniform_coeffs}
\end{equation}
There are 4 correct solutions for this problem, which are  $\{0101, 0110, 1000, 1001\}$, so that $P_R=\frac{1}{4}$. We run the code for this problem with various random initial values for the angle parameters and a sufficient number of iterations. Simulation results are shown in Fig.\ref{FIG_kbody_uniform_coeffs}. The results show that a single-layer QAOA circuit is insufficient to produce good outcomes, as indicated by the histogram in (a), where non-correct solutions still have high probabilities. Accordingly, we increase the number of layers subsequently into $p=4, 8, 16,$ and $32$. The histogram in (b) shows the solution distribution for $p=32$, where some of the correct solutions exhibit dominant probability values, while the incorrect ones are suppressed. The curve in (c) shows that the error stabilizes and converges after approximately 500 iterations of updating the angle parameters. The table in (d) shows that increasing the number of QAOA layers $p$ consistently enhances performance, as reflected by a reduction in objective errors and an increase in xRAR values.

\begin{figure*}[h!]
    \centering
    \subfloat[\centering Distribution of solution for $p=1$]
    {{\includegraphics[width=0.45\columnwidth]{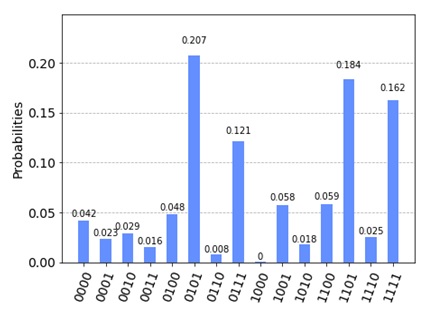} }} 
    \subfloat[\centering Distribution of solution for $p=32$]
    {{\includegraphics[width=0.45\columnwidth]{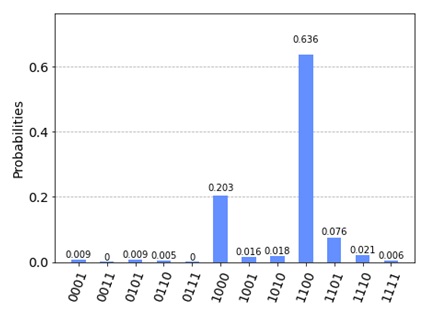} }} 
    \\
    \subfloat[\centering Iteration curve for $p=32$]
    {{\includegraphics[width=0.48\columnwidth,valign=t]{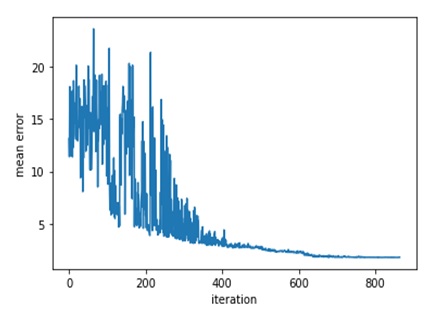} }}%
    \subfloat[\centering Performance Table]
    %
    {{
    \begin{tabular}[t]{|c|c|r|r|}
            \hline
            No & N-layer $p$ & Obj. Error&    xRAR\\
            \hline
            1&  1&   12.906&   0.938\\
            \hline
            2&  4&   10.049&   0.422\\
            \hline
            3&  8&   4.023&   5.766\\
            \hline
            4&  16&   5.275&   0.541\\
            \hline
            5&  32&   1.824&   10.172\\
            \hline
      \end{tabular}
      \vphantom{\includegraphics[width=0.48\columnwidth,valign=t]{img/k_bodies_non_uniform_coeffs_iter_p_2NQ_sqr.jpg}}
    }}
    \caption{Experimental results for a problem Hamiltonian with uniform $k$-body interactions but non-uniform coefficients: (a) The distribution of solutions for a single layer shows that while the correct solution has a high probability, other incorrect solutions also exhibit significantly high probabilities. (b) After increasing the layer number to 32, the distribution of solutions improves, with correct solutions now holding the highest probability. (c) The error curve indicates a decreasing error value as the iteration number increases. (d) The performance table by the number of layers demonstrates that increasing the layer count generally enhances xRAR performance and reduces errors; however, there are instances where performance may decrease.}%
    \label{FIG_kbody_non_uniform_coeffs}%
\end{figure*}

For the second case, we consider a more general Hamiltonian where the circuits of mixed $k$ body interactions is weighted with non-uniform coefficients as follows 
\begin{equation}
    \hat{H}\left( \hat{\sigma}\right)= \hat{\sigma}^z_{0} +2\hat{\sigma}^z_{1} + 3\hat{\sigma}^z_{2} + 5\hat{\sigma}^z_{0}\hat{\sigma}^z_{1} + 7\hat{\sigma}^z_{1}\hat{\sigma}^z_{2} + 11\hat{\sigma}^z_{1}\hat{\sigma}^z_{2}\hat{\sigma}^z_{3} + 13\hat{\sigma}^z_{0}\hat{\sigma}^z_{1}\hat{\sigma}^z_{2}\hat{\sigma}^z_{3}
\end{equation}
This problem only have one correct solution, which is $1100$, so that the probability of random algorithm is $P_R= 1/16$. Some of the simulation results are displayed in Fig.\ref{FIG_kbody_non_uniform_coeffs}. In parts (a) and (b), we observe a re-concentration of the distribution toward the solution as the number of layers $p$ increases. In (c), we observe that the error decreases and approaches convergence after approximately 800 iterations. The performance table in (d) indicates that increasing $p$ generally enhances performance by increasing xRAR and decreasing objective errors; however, at certain values of $p$, performance begins to degrade

%
In general, we can say that the performance of each individual circuit met our expectations, with the solution distributions confirming the circuit's validity. A more detailed results for other configurations are provided in the Supplementary Information.


%
\subsubsection{Finding Hadamard Matrices}
\input{2_3_1_HSEARCH_Williamson}

\input{2_3_2_HSEARCH_Turyn}

\input{2_3_3_HSEARCH_Higher_Order_Matrices}


%% file: 2_3_1_HSEARCH_Williamson.tex

The lowest order case for the Williamson method is 12, which corresponds to 36-order Baumert-Hall Hadmard matrix, requires $N_Q=8$ qubits to implement. The energy function and the Hamiltonian of this problems was obtained similarly to our previous paper \cite{suksmono_2022}. 

\begin{figure}[h!]
    \centering
    \includegraphics[width=1.0\columnwidth] {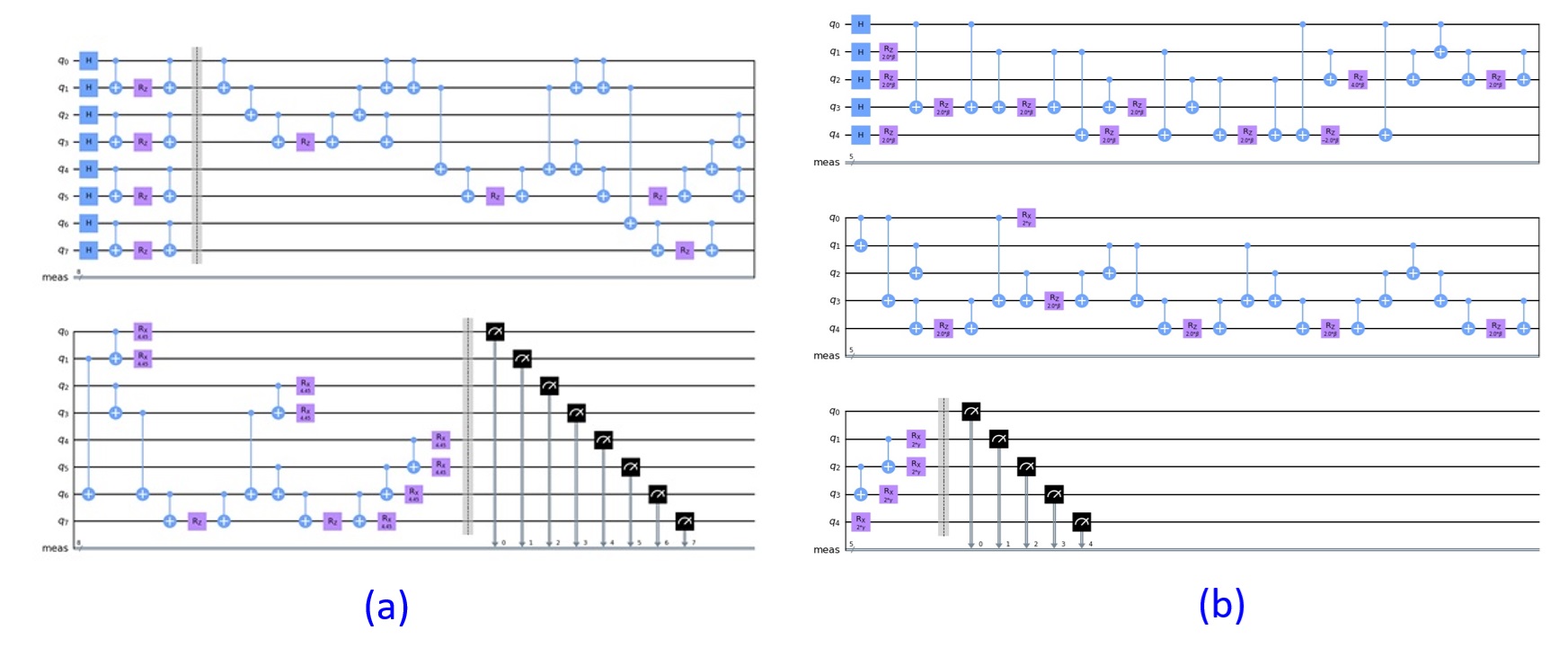} 
    \caption{Quantum Circuits of QAOA-Based Hadamard Search: (a) A 1-layer quantum for 12-order QAOA-Williamson/Baumert-Hall method, and (b) A 1-layer quantum circuit of 44-order QAOA-Turyn method.}
    \label{FIG_QC_WH12_TH44}
\end{figure}

An exhaustive search to all possible $2^{8}$ bit strings that yields minimum energy; and therefore correct bit strings, found $64$ WBH-sequences as correct solutions. This result yields the probability value to find the solution of 8-length uniformly distributed bit strings $P_R=\frac{1}{4}$; therefore, the maximum xRAR performance is $\frac{1}{P_R} = 4$. It was also found that the maximum value of the cost function is $18$ and the total error of $1,024$; implying that the normalized average error is equal to $0.2222$. The Hamiltonian of this problem is given by 
\begin{equation}
  \begin{split}
    \hat{H}_C\left( \hat{\sigma}\right)=2 \hat{\sigma}^z_0\hat{\sigma}^z_1 + 2\hat{\sigma}^z_2\hat{\sigma}^z_13 + 2\hat{\sigma}^z_4\hat{\sigma}^z_5 + 2\hat{\sigma}^z_6\hat{\sigma}^z_7 
    + \hat{\sigma}^z_0\hat{\sigma}^z_1\hat{\sigma}^z_2\hat{\sigma}^z_3 + \hat{\sigma}^z_0\hat{\sigma}^z_1\hat{\sigma}^z_4\hat{\sigma}^z_5 + \\ \hat{\sigma}^z_0\hat{\sigma}^z_1\hat{\sigma}^z_6\hat{\sigma}^z_7 
    + \hat{\sigma}^z_2\hat{\sigma}^z_3\hat{\sigma}^z_4\hat{\sigma}^z_5 + \hat{\sigma}^z_2\hat{\sigma}^z_3\hat{\sigma}^z_6\hat{\sigma}^z_7 +  \hat{\sigma}^z_4\hat{\sigma}^z_5\hat{\sigma}^z_6\hat{\sigma}^z_7 
    + 4
  \end{split}
  \label{EQ_H_williamson_12}
\end{equation}
Note that in the minimization, the constant term can be dropped without affecting the result. We will perform some experiments for this case with both of the simulator and the real quantum computer.

\begin{figure*}[t]
    \centering
    {{\includegraphics[width=0.90\columnwidth]
    {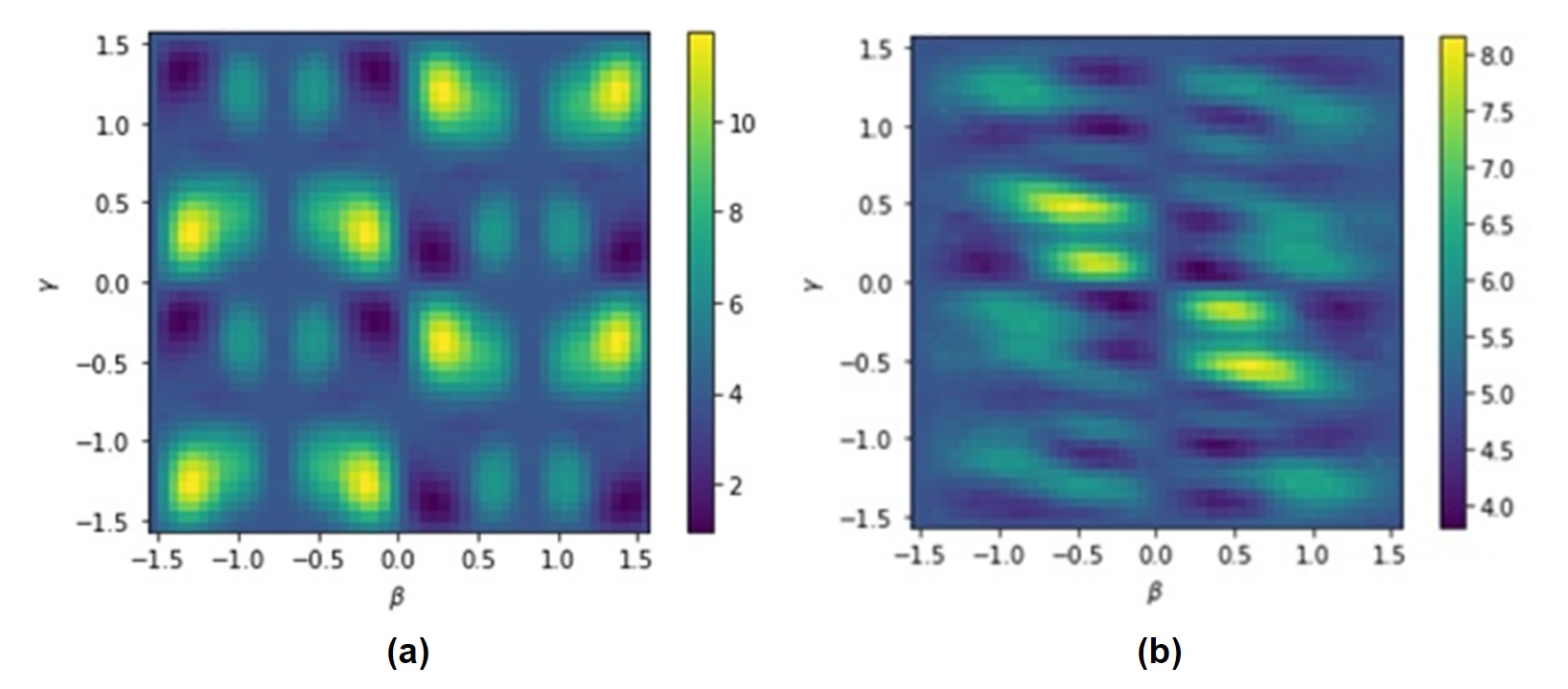} }} %
    \caption{Lowest order PEL (Potential Energy Landscape) of (a) 12-order QAOA-Based Williamson Method and (b) 44-order QAOA-Based Turyn Method. The Williamson's PEL method is relatively more regular than the Turyn's, indicating that finding the minimum value in the Turyn-based method is more difficult than the Williamson's.}%
    \label{FIG_WBH_QC}%
\end{figure*}

First, we run the QAOA-HSEARCH algorithm in a quantum computer simulator (IBM-Qiskit) with various number of layers, random initialization of $\{\beta_0, \gamma_0\}$ parameters, and apply the COBYLA (Constrained Optimization BY Linear Approximation) optimization method which is available in the Python library \cite{powell_2012}. Figure \ref{FIG_QC_WH12_TH44} (a) depicts a one-layer quantum circuit associated with the QAOA algorithm, with the Problem Hamiltonian given in Eq. (\ref{EQ_H_williamson_12}).

\begin{figure*}[t]
    \centering
    \subfloat[\centering Quantum Simulator]
    {{\includegraphics[width=0.50\columnwidth]{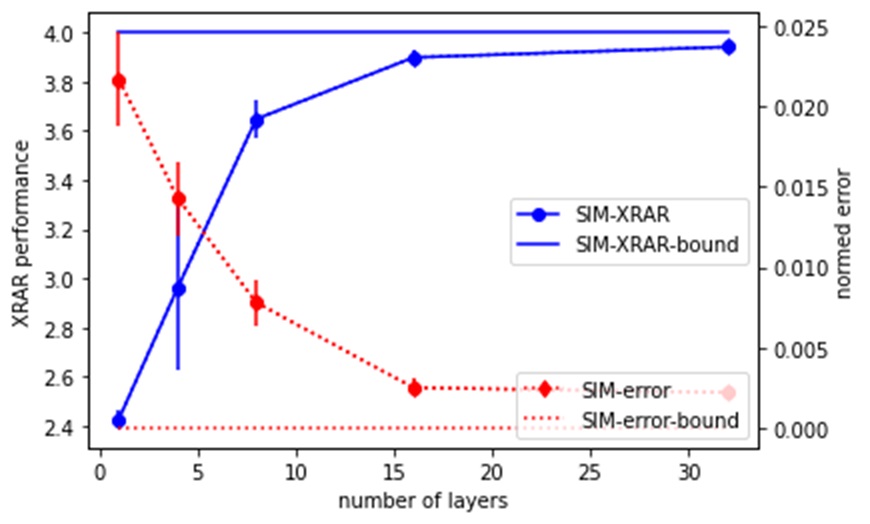} }} %
    \subfloat[\centering Quantum Hardware]
    {{\includegraphics[width=0.50\columnwidth]{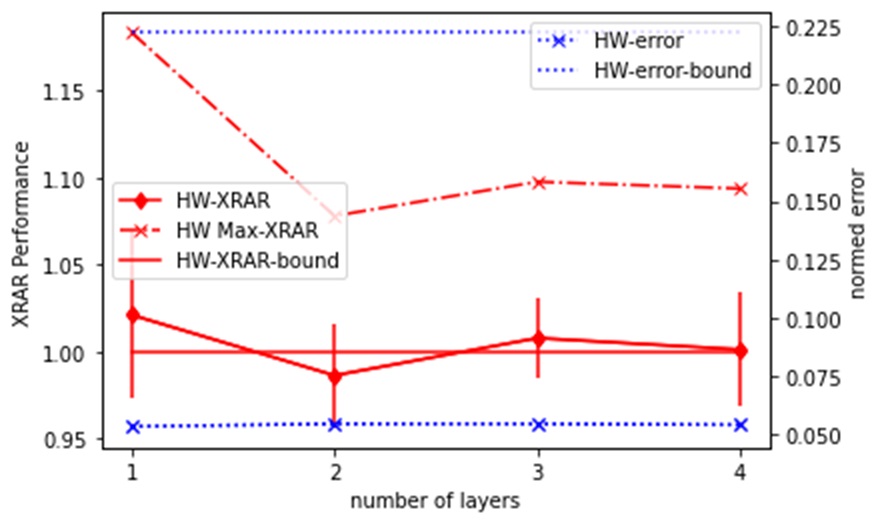} }} %
    \caption{Performance of 12-order Williamson/36-order Baumert Hall Based QAOA Methods. Figure (a) presents the simulation results on a quantum simulator: solid blue line with blue circles is the xRAR curve, solid blue line is the upper bound of xRAR value; which is equal to 4, the red-dotted line with circle is the normalized objective error, and the dotted red line is the lower bound of error, which is equal to zero. Fig. (b) Shows the hardware performance: the dotted blue line with $\times$ symbols at the bottom part is the curve of the objective error, the dotted blue line at the upper part is the error threshold for the random algorithm, solid red line with circle is the mean value of xRAR, the solid red line is the xRAR of random algorithm, and the red dashed-dot with $\times$ symbols are maximum value of xRAR at corresponding layer number.}%
    \label{FIG_WBH}%
\end{figure*}

\begin{figure*}[h!]
    \centering
    \subfloat[\centering 12-order Williamson]
    {{\includegraphics[width=0.3\columnwidth]{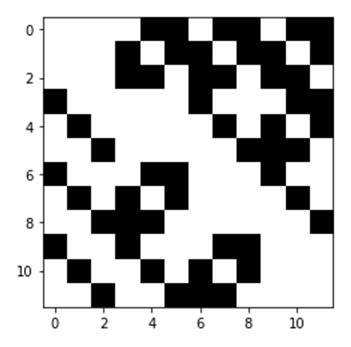} }}%
    \hspace{2 cm}
    \subfloat[\centering 36-order Baumert-Hall]
    {{\includegraphics[width=0.3\columnwidth]{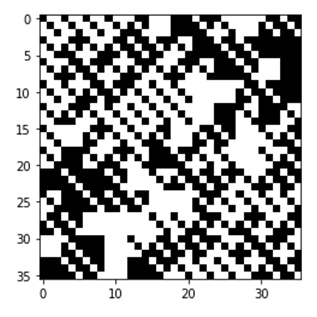} }}%
    \caption{The 12-order Williamson and 36-order Baumert-Hall Hadamard Matrices which are found by the proposed methods. Both execution of the algorithm in the quantum simulator and quantum hardware found identical H-matrices. In the figure, white boxes represent "+1" elements and the black ones represent "-1" elements of the matrices.}%
    \label{FIG_WBH_mtx}%
\end{figure*}

The energy distribution as a function of $\gamma$ and $\beta$ parameters displayed as PEL (Potential Energy Landscape) in Fig.\ref{FIG_WBH_QC} (a) reveals a periodic landscape with minima located in the first (right upper) and third (left lower part) quadrants. Additionally, the PEL for the 44th-order Turyn Hadamard search is shown in Fig. \ref{FIG_WBH_QC}(b) and will be discussed later. Fig.\ref{FIG_WBH} (a) shows the performance of the algorithm with the number of layers $p$ are increased stepped wisely, i.e, $p= 1, 4, 8, 16, 32$. Based on the location of the minima indicated in the PEL, the initialization of the parameters was selected within the interval $(-0.5, 0.5)$. We repeat the experiment 10 times and plot the mean value of xRAR and the error values in the figure. We observed that the value of xRAR consistently increased asymptotically to its maximum theoretical value of $xRAR=4$ at $p=32$. At the same time, we observed that increasing the number of layer reduces the error. The resulting 12-order of the Williamson's and its corresponding 36 order of Baumert-Hall's are displayed in Fig. \ref{FIG_WBH_mtx} (a) and Fig. \ref{FIG_WBH_mtx} (b), respectively.  

We also implemented the algorithm of finding 12-order Williamson matrix in a quantum computer hardware.  An IBM quantum computer, in this case is the IBM-Brisbane machine powered by a 127 qubits Eagle r.3 of version 1.1.6 quantum processor, was employed. The processor's qubits mean coherence time are $T_1 \approx 227 \mu s, T_2 \approx  130 \mu s$ with median ECR error $ \approx 7 \times 10^{-3}$ and median SX error  $ \approx 2 \times 10^{-4}$. We also repeat the run 10 times and the number of shots in the hardware is set to $1,024$. The results are displayed in Fig. \ref{FIG_WBH}(b), which is the quantum hardware (QPU) performance for the QAOA 12-order Williamson method with number of layers $1, 2, 3$ and $4$. This figure shows that, although the mean error of QAOA implemented on hardware (blue dotted line with "$\times$" symbols) are consistently lower than the mean error of random algorithm (blue dotted line in the upmost part), the mean xRAR performance (red solid line with red circle symbols) is sometimes slightly better than the random algorithm bound (red solid line) for number of layer of 1 and 3, and worse for 2 and 4. Since initialization of the angle can influence the final results, in term of xRAR, we also display the maximum xRAR for each repeated 10 times run. The max xRAR performance curve (red dashed-dot line with "$\times$" symbol) initially higher than random but then tend to decrease when the number of layers are increased. This demonstrates that increasing the circuit depth raises the noise level in the qubits, ultimately degrading the performance of our QAOA-based algorithm.

%

%% file: 2_3_2_HSEARCH_Turyn.tex
%
In the Turyn-based method, for a particular order of H-matrix that we want to construct, we have to find a corresponding TT (Turyn Type)-Sequence \cite{turyn_1974, kharaghani_rezaie_2004, london_2013, suksmono_2022}. In term of the energy function (see Eq.(\ref{EQ_cost_turyn}) in the Method section), we are looking for a T-string $\vec{s}$. For even positive integers $N=4, 6, 8, ... $, the order of related Turyn's Hadamard matrix will be $M=4(3N-1)$ and the number of variables or required qubits is $Q=4N-11$. We performed experiments for $N=4, 6, 8$ that corresponds to the Hadamard matrices of order $M=44, 68, 92$ which require $Q=5, 13, 21$ qubits, respectively.

In the first experiment, our goal is to find a Turyn H-matrix of order $44$, which needs 5 qubits to implement. The problem Hamiltonian is given by the following expression,
\begin{equation}
  \label{Hs_44_Turyn}
  \begin{split}
    \hat{H}_C\left( \hat{\sigma}\right)=\hat{\sigma}^z_0\hat{\sigma}^z_1\hat{\sigma}^z_2 + \hat{\sigma}^z_0\hat{\sigma}^z_3\hat{\sigma}^z_4 + \hat{\sigma}^z_0\hat{\sigma}^z_3 - \hat{\sigma}^z_0\hat{\sigma}^z_4 + \hat{\sigma}^z_1\hat{\sigma}^z_2\hat{\sigma}^z_3\hat{\sigma}^z_4 \\
    + \hat{\sigma}^z_1\hat{\sigma}^z_2\hat{\sigma}^z_3 + 2\hat{\sigma}^z_1\hat{\sigma}^z_2 + \hat{\sigma}^z_1\hat{\sigma}^z_3\hat{\sigma}^z_4 + \hat{\sigma}^z_1\hat{\sigma}^z_3 + \hat{\sigma}^z_1\hat{\sigma}^z_4 + \hat{\sigma}^z_1 + \hat{\sigma}^z_2\hat{\sigma}^z_3\hat{\sigma}^z_4 \\
    + \hat{\sigma}^z_2\hat{\sigma}^z_3 + \hat{\sigma}^z_2\hat{\sigma}^z_4 + \hat{\sigma}^z_2 + \hat{\sigma}^z_4 + 5
    \end{split}
\end{equation}
The corresponding quantum circuit can be automatically generated using a construction algorithm, with a single-layer version shown in Fig. \ref{FIG_QC_WH12_TH44} (b). The PEL shown in Fig. \ref{FIG_WBH_QC}(b) exhibits a more irregular surface compared to the 12th-order Williamson PEL in Fig.\ref{FIG_WBH_QC}(a). Therefore, we anticipate that finding a solution will be more challenging in the Turyn case than in the Williamson case. 

We then execute the QAOA on both a quantum simulator and quantum hardware, successfully identifying the 44th-order Turyn H-matrix. As illustrated in the QAOA flowchart in Fig. \ref{FIG_experiment_flow}, the quantum computing process is typically embedded within the optimization loop, requiring dedicated access to the quantum device. However, since we used a public access to a 5 qubits IBM quantum computer, such dedicated access was not permitted. Therefore, we only implemented the optimized quantum circuits obtained in the simulation into the 5 qubits IBM Quito.

\begin{figure*}[h!]
    \centering
    \subfloat[\centering Solution histogram produced by IBM Quito]
    {{\includegraphics[width=0.6\columnwidth]{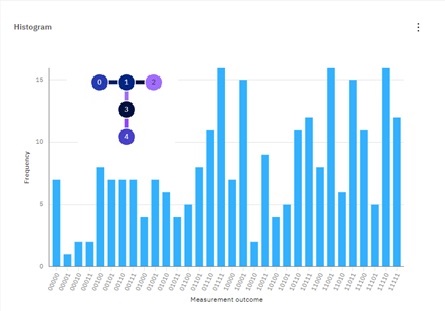} }}%
    \subfloat[\centering 44-order Turyn H-matrix]
    {{\includegraphics[width=0.35\columnwidth]{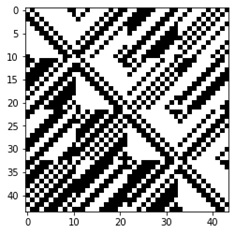} }}%
    \caption{Results for 44-order Turyn method: (a) Histogram of the output when running the algorithm in IBM Quito. The inset shows qubit configuration on the quantum device. (b) A Turyn Hadamard matrix of order-44 found by the proposed method running on IBM Quito}%
    \label{FIG_turyn_44}%
\end{figure*}

The resulting 44-order H matrix is identical for both the simulator and the hardware, which is shown in Fig. \ref{FIG_turyn_44}: (a) output histogram of implemented algorithm in IBM-Quito, and (b) result of 44-order of the Turyn H-matrix. The histogram in Fig. \ref{FIG_turyn_44} (a) indicates that the number of the valid solution, i.e. the bit string 11100, is equal to 11; which means that it consists of about $6 \%$ valid solution. Since a random algorithm would yield only $3.1\%$, the experiments on the quantum processor suggest nearly doubling the advantage. 

%% file: 2_3_3_HSEARCH_Higher_Order_Matrices.tex

\begin{figure*}[h!]
    \centering
    \subfloat[\centering 44-order Williamson Hadamard matrix]
    {{\includegraphics[width=0.45\columnwidth]{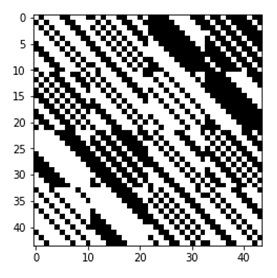} }}%
    \subfloat[\centering 132-order Baumert-Hall Hadamard matrix]
    {{\includegraphics[width=0.45\columnwidth]{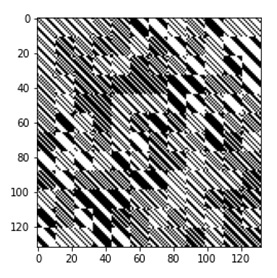} }}%
    \caption{Experiment Results QAOA-Based Hadamard Matrix Search on a Simulator: (a) Williamson Matrix of Order 44 and (b) Corresponding Baumert-Hall of Order 132}%
    \label{FIG_WBH_44_132}%
\end{figure*}

The PEL of the QAOA for higher-order H matrices is more irregular compared to that of lower-order matrices. This suggests that selecting initial parameters is both challenging and crucial. In the final experiment, we present the results of finding Williamson matrices of order 44 and Baumert-Hall matrices of order 132 using a quantum computer simulator. We used a single-layer QAOA and experimented with various initial parameters, selecting the best-performing configuration. The required number of qubits to implement this scheme is 24. After setting the number of sampling to 10,000 shots, we obtained the mean xRAR on 10 different parameter initialization is equal to 1.14 with the maximum value of 3.50. One of the obtained matrix is displayed in Fig. \ref{FIG_WBH_44_132}, where (a) shows 44-order Williamson and (b) its corresponding 132-order Baumert-Hall matrices.

%% file: 3__Discussions.tex
\section{Discussion}
We have demonstrated the feasibility of implementing Hadamard matrix search algorithms on a circuit-based quantum computer, both in a simulator and on actual quantum hardware. Within the framework of quantum optimization, we utilized the QAOA to construct the Hamiltonian described in our previous work \cite{suksmono_2022}. This Hamiltonian was then implemented in quantum circuits and executed on circuit-based quantum computers. Due to hardware limitations and the current state of noisy qubits, the implementation on quantum hardware was limited to the lowest-order cases with a single-layer QAOA scheme. However, the quantum simulator successfully handled higher-order cases.

Experimental results suggest that the difficulty in finding Hadamard matrices using QAOA algorithms arises from the non-smoothness of the energy landscape (PEL), which becomes increasingly pronounced in higher-order cases. While the Turyn-based method is more qubit-efficient than the Williamson and Baumert-Hall (WBH) method, its more irregular energy landscape makes finding Turyn's solution more challenging.

Experiments with the lowest-order WBH case (as shown in Fig. \ref{FIG_WBH}) on the quantum simulator demonstrate that increasing the number of layers consistently enhances performance. This improvement is indicated by the xRAR metric, which asymptotically approaches the performance limit as the number of layers increases. However, implementing the algorithm on a real quantum device did not replicate this performance. With a single layer, the algorithm performed slightly better than a random algorithm on average, but this advantage diminished as the number of layers increased—unless only when the best performance from multiple iterations was selected. However, the advantage of using more than one layer also disappears. These results suggest that increasing the number of layers on a NISQ device provides only limited benefits. This confirms the findings of previous QAOA researchers, including Boulebnane et al. \cite{boulebnane_2023}.

In late 2023, IBM successfully built a 1,121-qubit processor, known as the Condor quantum processor, although the issue of noise remains unresolved. But more recently, quantum error correction experiments have reached the threshold for the surface code \cite{acharya_2024}. If these trends continue, it is likely that some of the currently unknown Hadamard matrices will eventually be discovered by quantum computing. The QAOA method would require 336 qubits to find the lowest unknown 668-order H-matrix using the Williamson method, or 157 qubits using the Turyn method. In terms of qubit numbers, this is within the reach of current technology. However, noise will remain a significant obstacle to implementation. Nonetheless, it would be exciting to further explore this domain, especially with dedicated access to a quantum device capable of running QAOA at full capacity.
%
%

%

%% file: 4__Methods.tex
\section{Methods}
\input{4_1_HSEARCH}

\input{4_2_Elementary_Circuits_Construction}

%% file: 4_1_HSEARCH.tex
\subsection{Finding H-Matrices as a Binary Optimization Problem}

A direct method to find an $M$-order a H-matrix, i.e. a binary orthogonal matrix of size $M\times M$, can be done by checking the orthogonality condition of all possible binary matrices $B=[b_{m,n}]$, where $b_{m,n} \in \{-1, +1\}$. The orthogonality test can be formulated as a cost function $C_D(B)$, which is the sum of the squared off-diagonal elements of an indicator matrix $D=[d_{m,n}]=B^TB$, which can be expressed by, 
\begin{equation}
    C_D(B) = C(b_{m,n}) = \sum_{m=0}^{M-1} \sum_{n=0}^{M-1} \left(d_{m,n}- MI_{m,n} \right)^2  
\end{equation}
where $I$ is an $M\times M$ identity matrix. When $C_D(B)=0$, then the matrix $B$ is orthogonal and therefore it is a H-matrix; otherwise it is not. This is not an efficient method because the number of binary matrices to check is $2^{M\times M}$. 

A more efficient way of finding the H-matrix is by employing the Williamson/Baumert-Hall \cite{hedayat_1978} or the Turyn methods \cite{turyn_1974, kharaghani_rezaie_2004, london_2013}. We also have developed optimization based methods that employs quantum computers to find the H-matrix, which are the QA (Quantum Annealing) direct method by representing each entries as a binary variable \cite{suksmono_minato_2019}, the QA Williamson/Baumert-Hall method, and the QA Turyn method \cite{suksmono_2022}. Whereas the number of variables in the QA direct method grows with the order $M$ by $O(M\times M)$, the QA Williamson/Baumert-Hall and the QA Turyn methods only grows by $O(M)$, which is more efficient in term of the number of the variables. However, when it is implemented on the present day quantum annealer, such as the D-Wave, not only each variable should be represented by a qubit, but additional \emph{ancillary} qubits are also required for representing  3-body and 4-body terms. Accordingly, the required number of qubits grows with the order of the matrix by $O(M\times M)$. Since the qubit is one of the most valuable resources in quantum computing, a more efficient method that can reduce the number of qubits is highly desired.

In the Williamson based method \cite{suksmono_2022}, we seek for a binary $\{-1, +1\}$ vector
\begin{equation}
   \vec{s}=\left[s_0, s_1, \cdots, s_n, \, \cdots, s_{N-1} \right] 
   \label{EQ_WBH_string}
\end{equation}
where $s_n \in \{-1,+1\}$, that minimize a Williamson cost function $C_W(\vec{s})$ that is given by,
\begin{equation}
    C_W(\vec{s}) = \sum_{i=0}^{K-1} \sum_{j=0}^{K-1} \left(v_{i,j}(\vec{s})- 4k\delta_{i,j}\right)^2
    \label{EQ_cost_williamson}
\end{equation}
In this equation, $v_{i,j}(\vec{s})$ is the elements of matrix $V$ that is constructed from four sub-matrices $A,B,C, $ and $D$ of dimension $K \times K$; that is,
\begin{equation}
    V = A^TA + B^TB + C^TC +D^TD    
\end{equation}
where $V=V(\vec{s}), A=A(\vec{s}), B=B(\vec{s}), C=C(\vec{s}), D=D(\vec{s})$ are sub-matrices whose elements include some particular elements of the vector $\vec{s}$. When $C_W(\vec{s})=0$, then the matrix $H$ of size $4K \times 4K$ given by the following block matrix
\begin{equation}
    H = 
    \begin{pmatrix}
        A & B &  C & D \\
       -B & A & -D & C \\
       -C & D &  A &-B \\
       -D &-C &  B & A
    \end{pmatrix}
\end{equation}
is Hadamard \cite{hedayat_1978}. A larger Baumert-Hall matrix can also be constructed from the same $\{A, B, C, D\}$ submatrices \cite{hedayat_1978}. We will call the binary representation of vector $\vec{s}$ given in Eq. (\ref{EQ_WBH_string}) that minimize Eq.(\ref{EQ_cost_williamson}) as a Williamson/Baumert-Hall string or a \emph{WBH-string}.

In the Turyn based method, we also seek for a vector $\vec{s}=[s_0, s_1, \cdots , s_{N-1}]$ like in Eq.(\ref{EQ_WBH_string}) that minimize a Turyn cost function $C_T(\vec{s})$ given by
\begin{equation}
    C_T(\vec{s}) = \sum_{r>1} \left( N_{X(\vec{s})}(r) + N_{Y(\vec{s})}(r) + 2N_{Z(\vec{s})}(r) + 2N_{W(\vec{s})}(r) \right)^2
    \label{EQ_cost_turyn}
\end{equation}
where $N_{X(\vec{s})}(r), N_{Y(\vec{s})}(r), N_{Z(\vec{s})}(r), N_{W(\vec{s})}(r)$ are non-periodic auto-correlation functions of sequences $X(\vec{s}), Y(\vec{s}), Z(\vec{s}), W(\vec{s})$, respectively, which are calculated at lagged $r$. Note that for a sequence $X=[x_0, x_1, \cdots, x_{N-1}]$, the non-periodic auto-correlation function is given by \cite{kharaghani_rezaie_2004, london_2013},

\begin{equation}
N_X(r) = 
     \begin{cases}
       \sum_{n=0}^{N-1-r} x_n x_{n+r} &, 0 \leq r \leq N-1\\
       0 &, r \ge N\\
     \end{cases}
\end{equation}
Similarly as in the previous case, we will call the binary representation of vector $\vec{s}$ that makes $C_T(\vec{s})=0$ as a Turyn string or \emph{T-string}. In this paper, since the number of variables can be very large, the computation of the cost functions $C_W(\vec{s})$ and $C_T(\vec{s})$ and its corresponding Hamiltonian expression are performed by symbolic computing. These functions are then transformed into the problem Hamiltonian $\hat{H}_C$ in Eq.(\ref{EQ_HC_HB}).

%% file: 4_2_Elementary_Circuits_Construction.tex
\subsection{Construction of Elementary Circuits}

Consider a general problem Hamiltonian given by Eq. (\ref{EQ_QAOA_problem_hsearch}). By using Eq.(\ref{EQ_QAOA_general}), the unitary for a single layer problem's Hamiltonian can be expressed by
\begin{equation}
    \label{EQ_problem_h_unitary}
   \hat{U}(\gamma)=\prod_{j}e^{-i\gamma c_j\hat{\sigma}^z_j}
            \prod_{j,k}e^{-i\gamma c_{jk}\hat{\sigma}^z_j\hat{\sigma}^z_k}
            \prod_{j,k,m}e^{-i\gamma c_{jkm}\hat{\sigma}^z_j\hat{\sigma}^z_k\hat{\sigma}^z_m}
            \prod_{j,k,m,n}e^{-i\gamma c_{jkmn}\hat{\sigma}^z_j\hat{\sigma}^z_k\hat{\sigma}^z_m\hat{\sigma}^z_n}
\end{equation}
We can represent the exponentiation of $\hat{\sigma}^z$ as a rotation in $z$-direction, $\hat{R}_Z(\cdots)$, as  follows
\begin{equation}
    \nonumber
     \hat{U}(\gamma) = e^{-i\gamma \hat{\sigma}^z} 
     = e^{-i\gamma 
     \begin{pmatrix}
      1 & 0 \\
      0 & -1
     \end{pmatrix}
     }
     =  
     \begin{pmatrix}
      e^{-i\gamma} & 0 \\
      0 & e^{i\gamma}
     \end{pmatrix}
     = \hat{R}_Z(2\gamma)
\end{equation}
Substituting $\gamma \leftarrow c_j \gamma$ yields
\begin{equation} 
    \nonumber
     \hat{U}(c_j\gamma)= e^{-c_j\gamma \hat{\sigma}^z} = \hat{R}_Z(2c_j\gamma)
\end{equation}
%

%
The first term of the product in Eq.(\ref{EQ_problem_h_unitary}), considering there are $N$ qubits to be rotated, can be expanded into the following

\begin{equation}
    \nonumber
    \prod_{j} e^{-ic_j\gamma \hat{\sigma}_j^z} =  
    \begin{pmatrix}
      e^{-ic_0\gamma} & 0 \\
      0 & e^{ic_0\gamma}
    \end{pmatrix}
    \otimes
    \begin{pmatrix}
      e^{-ic_1\gamma} & 0 \\
      0 & e^{ic_1\gamma}
    \end{pmatrix}
    \otimes
    \cdots
    \otimes
    \begin{pmatrix}
      e^{-ic_{N-1}\gamma} & 0 \\
      0 & e^{ic_{N-1}\gamma}
    \end{pmatrix}
\end{equation}
where $\otimes$ denotes tensor product. By denoting the $Z$-rotation on qubit $n$ as $\hat{R}_{Z_n(\cdots)}$, we can write
\begin{equation}
    \nonumber
    \prod_{j} e^{-ic_j\hat{\sigma}_j^z} 
    = \hat{R}_{Z_0}(2c_0\gamma) \otimes \hat{R}_{Z_1}(2c_1\gamma) \otimes \cdots \otimes \hat{R}_{Z_{N-1}}(2c_{N-1}\gamma)
    \label{EQ_unitary_1body_prod}
\end{equation}
%
The products in Eq.(\ref{EQ_problem_h_unitary}) can be interpreted as a series of cascaded operations. Then, a 1-body interaction term in the Hamiltonian that is expressed by $\hat{H}_{C_1}= \sum_j c_j \hat{\sigma}^z_j$ can be implemented as a quantum circuit given by Fig. \ref{FIG_1BODY}.(a), which is a Z-rotation of angle $2c_j\gamma$.
%
Higher order terms, which are 2-,3-, and 4- body interacting terms, can also be treated similarly, but with a different elementary circuits in the cascaded block. 

The quantum circuits implementation of the $k$-body interactions displayed in Fig. \ref{FIG_1BODY} (b), (c), and (d) are adopted from Nielsen-Chuang \cite{nielsen_chuang_2010}, Seeley \cite{seeley_2012}, and Setia \cite{setia_2018}. A 2-body interaction term in the Hamiltonian $\hat{H}_{C_2}= \sum_{j<k} c_{jk}\hat{\sigma}^z_j\hat{\sigma}^z_k$ has a corresponding circuits given by Fig. \ref{FIG_1BODY}.(b), which is a combination of CNOT and Z-rotation gate. The 3-body interaction term in the Hamiltonian which is expressed by $\hat{H}_{C_3}= \sum_{j<k<m} c_{jkm} \hat{\sigma}^z_j\hat{\sigma}^z_k\hat{\sigma}^z_m$ has a corresponding circuits given by Fig. \ref{FIG_1BODY}.(c), which is a combination of CNOT and Z-rotation gate acting on 3 qubits. Finally, a 4-body interaction term in the Hamiltonian  $\hat{H}_{C_4}= \sum_{j<k<m<n} c_{jkmn} \hat{\sigma}^z_j\hat{\sigma}^z_k\hat{\sigma}^z_m\hat{\sigma}^z_n$ has a corresponding circuits given by Fig.\ref{FIG_1BODY}.(d), which is a combination of CNOT and Z-rotation gate acting on 4 qubits.

%% file: 5_Acknowledgements.tex
\section*{Acknowledgments}
This work has been supported partially by the P2MI Program of STEI-ITB and by the Blueqat Inc., Tokyo, Japan.

\section*{Competing interests}
The authors declare no competing interests.

\section*{Author contributions statement}
A.B.S. formulated the theory, conducted the experiments, analyzed the results, and wrote the paper.

\section*{Data and Codes Availability}
All of codes and data will be provided upon direct request to the authors. Some parts of the codes will be made available for public upon publication of the manuscript.